\documentclass[conference]{IEEEtran}
\ifCLASSINFOpdf
\else
\fi
\usepackage{graphicx}
\usepackage{subfigure}
\begin{document}
%
\title{Theoretical Analysis of Cyclic Frequency Domain Noise and Feature Detection for Cognitive Radio Systems
}

\author{\IEEEauthorblockN{Gan Xiaoying\IEEEauthorrefmark{1},   Shen Da\IEEEauthorrefmark{1},   Zhou Yuan\IEEEauthorrefmark{2}, Zhang Wei\IEEEauthorrefmark{2}, Qian Liang\IEEEauthorrefmark{1}}
\IEEEauthorblockA{\IEEEauthorrefmark{1}Dept. of Electronic Engineering, Shanghai Jiao Tong University, Shanghai, 200240, China\\
Email: ganxiaoying@sjtu.edu.cn}
\IEEEauthorblockA{\IEEEauthorrefmark{2}Huawei Technologies Co.,
Ltd\\Email: micheal@huawei.com} }


%


\maketitle
\renewcommand{\thefootnote}{\fnsymbol{footnote}}

\begin{abstract}
In cognitive radio systems, cyclostationary feature detection plays
an important role in spectrum sensing, especially in low SNR cases.
To configure the detection threshold under a certain noise level and
a pre-set miss detection probability $P_f$, it's important to derive
the theoretical distribution of the observation variable. In this
paper, noise distribution in cyclic frequency domain has been
studied and Generalized Extreme Value (GEV) distribution is found to
be a precise match. Maximum likelihood estimation is applied to
estimate the parameters of GEV. Monte Carlo simulation has been
carried out to show that the simulated ROC curve is coincided with
the theoretical ROC curve, which proves the efficiency of the
theoretical distribution model.
\end{abstract}

\begin{IEEEkeywords} Cognitive radio, spectrum sensing, cyclic feature detection, noise distribution on cyclic frequency domain
\end{IEEEkeywords}

%
\IEEEpeerreviewmaketitle

\section{Introduction}
Spectrum sensing plays an important role in cognitive radio (CR) systems, and cyclostationary feature detection is one of the main technologies for
spectrum sensing in low SNR cases [1-2]. The idea of cyclic detection is that one CR node samples the RF signals, transform the time domain signals into
cyclic frequency domain, then decide primary user's occupancy of a target band regarding whether the cyclic spectrum on a significant cyclic frequency is
above a certain threshold [2].\\
\indent Although the simulation results of cyclostationary feature
detection has been studied [3-5], less attention has been paid to
analyze the noise distribution on cyclic frequency domain.
Consequently, theoretical function between detection threshold
$\lambda_{th}$ and miss detection probability $P_f$ is not
available, which makes it difficult for practical system design. For
comparison, the theoretical function between detection threshold
$\lambda_{th}$ and miss detection probability $P_f$ of energy detection
is available by using central and non-central chi-square distribution to
model the distribution of the observation variable affected by time domain
Gaussian noise [6], which makes energy detection to be a practical method
 for spectrum sensing. However, its performance in low SNR cases is much
 poorer than cyclostationary feature detection [7]. In this paper, noise
 distribution on cyclic frequency domain has been analyzed and
 generalized extreme value distribution is found to be a precise match of the observation
 variable affected by time domain Gaussian noise. A fast cyclic frequency
 domain feature detection algorithm [7] has been introduced to evaluate
 the coincidence between theoretical ROC curve and the simulated ROC curve,
 which proves the reliability of theoretical distribution model and feasibility
 of practical system design.  \\
\indent The rest part of the paper is organized as follows: Section
II describes the system model of cyclostationary feature detection.
Noise distribution on cyclic frequency domain is analyzed in section
III. A fast cyclic frequency domain feature detection algorithm has
been introduced in section IV. Simulation results are given in Section V.
Finally, conclusions are drawn in Section VI.\\

\section{ SYSTEM MODEL}
The spectrum sensing problem can be modeled as hypothesis testing.
It is equivalent to distinguishing between the following two
hypotheses:
\begin{equation}
\left\{ \begin{array}{l}
 H_0: y(t)=n(t)\\
H_1: y(t)=x(t)+n(t)\\
 \end{array} \right.
\end{equation}
$y(t)$, $x(t)$ and $n(t)$ denote the received signal, the primary
user's transmit signal and the Gaussian noise, respectively. $H_1$
and $H_0$ represent the hypothesis that the primary user is active
or inactive. Due to the existence of noise, a certain threshold
$\lambda_{th}$ should be set to decide whether a primary user is
active or not. Probability of detection ($P_d$) and false alarm
($P_f$) are defined to evaluate the detection performance:
\begin{equation}
\left\{ \begin{array}{l}
P_d=p(y_i>\lambda_{th}|H_1)\\
P_f=p(y_i>\lambda{th}|H_0)\\
 \end{array} \right.
\end{equation}
The goal of detection is to maximize $P_d$ while maintain a given
$P_f$.\\
\indent When feature detection is applied, the detection model (1)
changes into:
\begin{equation}
\left\{ \begin{array}{l}
 H_0: S_y^\alpha(f)=S_n^\alpha(f)\\
H_1: S_y^\alpha(f)=S_x^\alpha(f)+S_n^\alpha(f)\\
 \end{array} \right.
\end{equation}
$S_y^\alpha(f)$ is the spectrum correlation density (SCD) of the
received signal $y(t)$, $S_x^\alpha(f)$ and $S_n^\alpha(f)$ is the
SCD of $x(t)$ and $n(t)$, respectively [8].

\indent Theoretically, Gaussian noise $n(t)$ is not a
cyclostationary statistic process, then $S_n^\alpha(f)=0$ when
$\alpha\neq0$ [8]. As for cyclostationary signal $x(t)$, there is a
significant frequency set \{$\alpha_0$\}, on which
$S_x^{\alpha_0}(f)\neq0$. Due to the ideal non-cyclostationary
characteristic of Gaussian noise, any pre-set threshold
$\lambda_{th}$ on a significant frequency will lead to $P_d=1$ and
$P_f=0$.\\

\section{Noise distribution on cyclic frequency domain}
In practice, SCD is calculated for limited length signals,
therefore, $S_n^\alpha(f)\neq0$ when $\alpha\neq0$ [9]. As shown in
Fig.1, the background noise is obvious on $f\sim\alpha$ square when
calculating SCD of a noise interfered AM modulated signal.
\begin{figure}[htbp] \centering
\includegraphics[width=0.45\textwidth]{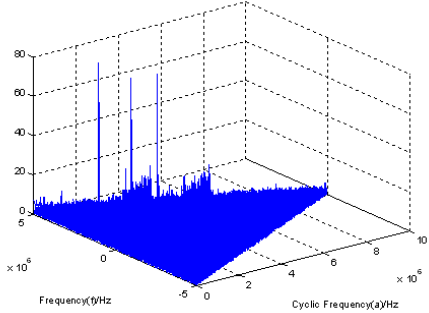}
\caption{Cyclic Spectrum Density of AM at SNR=-10db}\label{fig_sim}
\end{figure}

\indent In order to analyze the background noise distribution, a
limited-length Gaussian noise sequence $n^{iK}(j)$ is considered:
\begin{equation}
n^{iK}(j)=\left\{ \begin{array}{l}
 n(iK+j),0\leq j \leq K-1\\
 0,else\\
 \end{array} \right.
\end{equation}
where $K$ is the length of the analysis window, $i=0,1,...,\infty$
is the index of analysis window. Noise data in each analysis window
are transformed into cyclic frequency domain, and then mapped from
$f\sim\alpha$ square to $\alpha$ axis through the following
expression:
\begin{equation}
N^{iK}(\alpha)=\max_{f}|S_n^\alpha(f)|
\end{equation}
\indent For each cycle frequency $\alpha_0$, the cyclic spectrum
value is aligned to a set \{$N_i=N^{iK}(\alpha_0)$\}, $i=0,1,...,L$.
According to the extreme definition of $N^{iK}(\alpha_0)$ in (5),
Generalized Extreme Value (GEV) distribution is adopted to model the
cyclic
frequency domain noise [10]. The density function is:\\
when $\kappa\neq 0$
\begin{eqnarray}
f(x|\kappa,\mu,\sigma)= (\frac{1}{\sigma})e^{(-(1+\kappa
\frac{(x-\mu)}{\sigma})^{-\frac{1}{\kappa}})}(1+\kappa
\frac{(x-\mu)}{\sigma})^{-1-\frac{1}{\kappa}}
\end{eqnarray}
when $\kappa=0$
\begin{equation}
f(x|0,\mu,\sigma)=(\frac{1}{\sigma})exp(-exp(-\frac{x-\mu}{\sigma})-(-\frac{x-\mu}{\sigma}))
\end{equation}
where $\kappa$ is the shape parameter, $\mu$ is the position
parameter, $\sigma>0$ is the scale parameter. And the parameters
$\kappa$ , $\mu$ , $\sigma$ can be estimated by maximum likelihood
estimation based on the noise sequence \{$N_i$\}. For most cases,
$\kappa\approx0$, then the likelihood function is defined as
follows:
\begin{equation}
l(\mu,\sigma)=-Lln\sigma-\sum_{i=1}^L(\frac{N_i-\mu}{\sigma})-\sum_{i=1}^Lexp\{-(\frac{N_i-\mu}{\sigma})\}
\end{equation}
Let $\frac{\partial l}{\partial \mu}=0$,$\frac{\partial l}{\partial
\sigma}=0$ then:
\begin{equation}
\left\{ \begin{array}{l}
 \sum_{i=1}^Le^{-(N_i-\hat{\mu})}/\hat{\sigma}=L\\
 \sum_{i=1}^L(N_i-\hat{\mu})(1-e^{-(N_i-\hat{\mu})/\hat{\sigma}})=L\hat{\sigma}\\
 \end{array} \right.
\end{equation}
where $\hat{\mu}$ and $\hat{\sigma}$ are the estimated value of
$\mu$ and $\sigma$. By solving (9), $\hat{\mu}$ and $\hat{\sigma}$
are obtained.\\
 \indent After that, the likelihood function for
$\kappa$ is defined as: \setlength\arraycolsep{2pt}
\begin{eqnarray}
l(\kappa,\hat{\mu},\hat{\sigma})&=&
-Nln\hat{\sigma}-(1+\frac{1}{\kappa})\sum_{i=1}^Nln[1+\kappa(\frac{n_i-\hat{\mu}}{\hat{\sigma}})]\nonumber\\
&&-\sum_{i=1}^N[1+\kappa(\frac{n_i-\hat{u}}{\hat{\sigma}})]^{-\frac{1}{\kappa}}
\end{eqnarray}
Let $\frac{\partial l}{\partial \kappa}=0$, the estimation of
$\hat{\kappa}$ can be obtained by solving (10).\\

\section{Fast Cyclic Frequency Domain Feature Detection Algorithm}
Fig.2 shows the block diagram of a cyclic frequency domain feature
detector [7]:
\begin{figure}[htbp]
\centering
\includegraphics[width=0.55\textwidth]{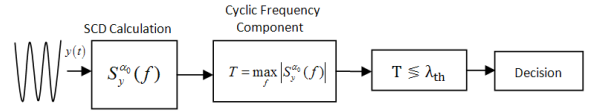}
\caption{Block diagram of a feature detector}\label{fig_sim}
\end{figure}

\indent In this detector, time domain signals are transformed to
cyclic frequency domain, and then mapped to $\alpha$ axis and get
extreme value for each $\alpha$, finally, compared with a pre-set
threshold $\lambda_{th}$ to determine the occupancy of primary user.
To decrease the computational complexity of feature detection, only
one cycle frequency $\alpha_0$ for a modulated signal is calculated
for spectrum sensing [7]. The probability of detection ($P_d$) and
false alarm ($P_f$) is defined by:
\begin{equation}
\left\{ \begin{array}{l}
P_d=Pr(T>\lambda_{th}|H_1)\\
P_f=Pr(T>\lambda{th}|H_0)\\
 \end{array} \right.
\end{equation}
where $Pr(\cdot)$ is CDF of generalized extreme value distribution.
Substitution of (6) and (7) into (11) yields:
\begin{eqnarray}
P_f=Pr(T>\lambda_{th}|H_0)=\int_{\lambda_{th}}^{+\infty}f(x|\hat{\kappa},\hat{\mu},\hat{\sigma})dx
\end{eqnarray}
For a pre-set $P_f$, the threshold can be estimated as:
\begin{equation}
\lambda_{th}=\left\{ \begin{array}{l}
\hat{\mu}-\frac{\hat{\sigma}}{\hat{\kappa}}(1-y_p^{}-\hat{\kappa}),\hat{\kappa}\neq0\\
\hat{\mu}-\hat{\sigma}logy_p,\hat{\kappa}=0\\
 \end{array} \right.
\end{equation}
where $y_p=-log(1-P_f)$.\\

\section{ Simulation Results}
In this section, Monte-Carlo simulation results are presented to
prove the reliability of the upper analytical results between $P_f$
and the threshold $\lambda_{th}$.\\
\indent Frequency smoothing method in [11-12] is applied to estimate
SCD of a time domain noise signal. Simulation parameters are listed
in TABLE 1. Length of the analysis window is set to be 4096 and
totally 10000 cyclic spectrum values on cyclic frequency
$\alpha_0=2f_c$, aligned by analysis window index, are considered to
evaluate the theoretical curve.
\begin{table}[htbp]
\renewcommand{\arraystretch}{1.5}
\caption{Simulation parameters list} \label{table_example}
\centering
\begin{tabular}{|c|c|}
\hline
Parameter & Value\\
\hline
Modulation type & AM\\
\hline
Carrier frequency &  1 MHz\\
\hline
  Bandwidth & 10 KHz\\
\hline
  Sampling frequency & 3 MHz\\
  \hline
   Sampling time & 1.365ms\\
  \hline
  Channel & AWGN\\
  \hline
  Window type & hamming\\
  \hline
   Frequency smoothing length & 1300\\
  \hline
  Sampled data length(K)& 4096\\
\hline
\end{tabular}
\end{table}
\indent The histogram of the aligned cyclic spectrum values is shown
in Fig 3, with compared to GEV distribution. It is proved that the
GEV distribution precisely match the cyclic frequency domain noise
data.
\begin{figure}[htbp]
\centering
\includegraphics[width=0.5\textwidth]{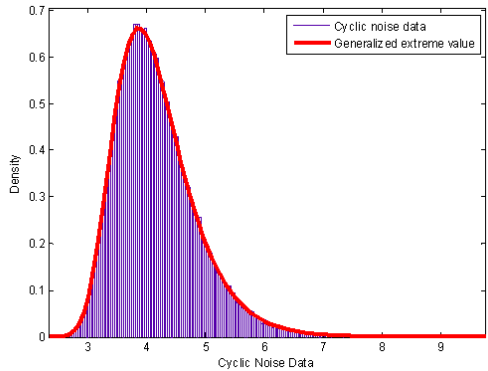}
\caption{GEV curve for the noise samples}\label{fig_sim}
\end{figure}\\
\indent For further proof of the proposed model, the curves of
\emph{receiver operating characteristics (ROC)} , which are
theoretically derived from GEV distribution, are plotted to compare
with those derived from Monte-Carlo simulation. As to the
theoretical curve, $P_f$ is pre-set according to system requirement.
By using (13), a theoretical threshold $\lambda_{th}$ is obtained.
Finally, received signals under hypothesis $H_1$ are compared with
the threshold to obtain the statistics result of $P_d$. As to the
simulated curve, threshold are chosen to be the same as the
theoretical threshold \{$\lambda_{th}$\}, after that, received
signals under hypothesis $H_1$ and $H_0$ are compared with each
$\lambda_{th}$ to obtain the statistics results of $P_d$ and $P_f$.
Finally, plot these
($P_d$,$P_f$) points to form a continuous curve.\\
\indent Experiments results are shown in Fig.4, we can see that the
theoretical ROC curve (red highlighted) precisely match the
simulated ROC curve (green highlighted) for different received
signal power levels. It is proved that the generalized extreme value
distribution is efficient to model the noise distribution on cyclic
frequency domain.
\begin{figure}[htbp]
\centering
\includegraphics[width=0.5\textwidth]{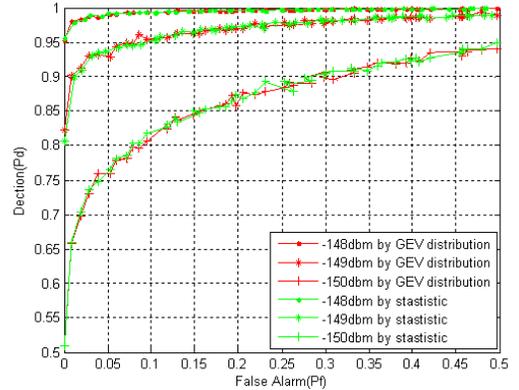}
\caption{ROC curves under different signal power
levels}\label{fig_sim}
\end{figure}

\section{Conclusion}
In this paper, noise distribution on cyclic frequency domain is
studied and generalized extreme value (GEV) distribution is found to
be an efficient method to model the cyclic frequency domain noise.
Maximum likelihood estimation is applied to estimate the parameters
of GEV. Sensing threshold is consequently derived from system
requirements (a pre-set $P_f$) and theoretical CDF of GEV
distribution. Monte Carlo simulation has been carried out to prove
that the simulated ROC curve is precisely coincided with the
theoretical ROC curve.


\section*{Acknowledgment}
The project is founded by the Corporation Research Department of
HUAWEI technology, the national 863 project of China, No.
2007AA01Z237, and the fund of Ministry of Science and Technology of
China, No. 2008DFA11950.



%

\end{document}